\documentclass[prl,floatfix,showpacs,twocolumn]{revtex4}

\usepackage{amsmath}
\usepackage{amssymb}
\usepackage{graphicx}
\usepackage[tight,sf]{subfigure}

\newcommand{\ie}{\emph{i.}$\,$\emph{e.}}

\newcommand{\etal}{\emph{et}$\,$\emph{al.}}
\newcommand{\cf}{\emph{cf.}\ }

\newcommand{\romd}{{\text{d}}}

\newcommand{\VECf}{{\boldsymbol{f}}}

\newcommand{\VECn}{{\boldsymbol{n}}}

\newcommand{\VECt}{{\boldsymbol{t}}}
\newcommand{\VECu}{{\boldsymbol{u}}}

\newcommand{\VECz}{{\boldsymbol{z}}}

\newcommand{\VECJ}{{\boldsymbol{J}}}

\newcommand{\VECX}{{\boldsymbol{X}}}

\newcommand{\VECphi}{{\boldsymbol{\varphi}}}

\newcommand{\CALE}{{\mathcal{E}}}

\newcommand{\CALK}{{\mathcal{K}}}
\newcommand{\CALO}{{\mathcal{O}}}

\newcommand{\ellipticK}[1]{\CALK ( #1 )}
\newcommand{\ellipticcE}[1]{\CALE ( #1 )}
\newcommand{\Jacobisnadjustroundb}[2]{\operatorname{sn}\left(  #1,#2 \right)}
\newcommand{\Jacobisnadjustsquareb}[2]{\operatorname{sn}\left[  #1,#2 \right]}
\newcommand{\Jacobiamadjust}[2]{\operatorname{am}\left(  #1,#2 \right)}

\newcommand{\RR}{\mathbb{R}}

\newcommand{\se}{\varphi_{\text{e}}}
\newcommand{\sek}{\varphi_{\text{e}}^{\text{kiss}}}
\newcommand{\sem}{\varphi_{\text{e}, \text{max}}^{\text{kiss}}}
\newcommand{\st}{s_{\text{e}}}
\newcommand{\sk}{s_{\text{kiss}}}

\newcommand{\FRAKL}{\mathcal{L}}

\newcommand{\econe}{\emph{e}-cone}
\newcommand{\econes}{\emph{e}-cones}

\usepackage{color}

\begin{document}

\title{Conical defects in growing sheets}

\author{Martin Michael M\"uller}
\author{Martine Ben Amar}
\affiliation{Laboratoire de Physique Statistique de l'Ecole Normale Sup\'erieure
             (UMR 8550), %
             associ\'e aux Universit\'es Paris 6 et Paris 7 et au CNRS;
             24, rue Lhomond, %
             75005 Paris, %
             France}

\author{Jemal Guven}
\affiliation{Instituto de Ciencias Nucleares, %
             Universidad Nacional Aut\'onoma de M\'exico, %
             Apdo.\ Postal 70-543, %
             04510 M\'exico D.F., %
             Mexico}

\date{\today}
\begin{abstract}
A growing or shrinking disc will adopt a conical shape, its intrinsic geometry characterized by a surplus angle $\se$ at the apex. If growth is slow, the cone will find its equilibrium. Whereas this is trivial if $\se\le 0$, the disc can fold into one of a discrete infinite number of states if $\se >0$. We construct these states in the regime where bending dominates, determine their energies and how stress is distributed in them. For each state a critical value of $\se$ is identified beyond which the cone touches itself. Before this occurs, all states are stable; the ground state has two-fold symmetry.
\end{abstract}

\maketitle


\bibliographystyle{plain}


Soft matter systems may display enormous complexity at the microscopic 
level \cite{Jones,RubinsteinColby}. However, on mesoscopic or larger scales of 
physical interest, the relevant degrees of freedom very often turn out to be 
purely geometrical. If one dimension is much smaller than the two others, an 
effective description in terms of a two-dimensional surface becomes appropriate. 
This is just as true for biological membranes \cite{Helfrich,Seifert97} as it is 
for inanimate matter \cite{Wittenreview}.

The equilibrium shape of the surface is often a minimum of bending energy. 
Typically however, one must take into account external forces, or constraints on the 
geometry: these may be \emph{global} as in the fluid membranes occurring in cells 
in which the area or the enclosed volume is fixed; they may also be \emph{local} 
as in plant tissues which are described, to a good approximation, as an 
unstretchable surface with a fixed metric \cite{Sharon,JulienMartine}. 

In general, bending is not possible without stretching. It turns out, however, 
that the most effective way to minimize the energy and satisfy the constraint 
is by confining the regions where stretching occurs to a series of sharp peaks and 
ridges \cite{Witten,benpom}. On length scales much larger than the thickness of 
the sheet, these can then be treated as points and curves along which boundary 
conditions are set on the surface. The simplest geometry of this kind is 
the developable cone--the ``point defect'' of folding, with the bending energy 
localized near the apex. Such a geometry is illustrated beautifully by a flat 
planar disc of paper depressed into a circular frame by applying a point force to 
its center \cite{cerdamaha}. 

Conical shapes also occur in living tissues. The unicellular algae 
\emph{Acetabularia acetabulum}, for example, grows a conical cap in the course of 
its development \cite{JulienMartine,algae}. Despite the superficial similarity, 
however, the point-like singularity exhibited in such cones is very different from the one which appears in the developable cone. In the latter, the singularity at the apex 
is extrinsic; the metric itself remains the same as that of the original disc. 
This is captured by the fact that the surplus angle at the apex vanishes: the length 
of the closed curve at a unit distance from the apex is equal to $2\pi$. If however 
there is a surplus or a deficit, there will be a non-trivial folded state even when 
external forces do not act. Whereas this state is an unremarkable circular cone in 
the case of a deficit, when the deficit is turned to surplus, the folded shape--an 
\emph{excess}-cone (\econe\ for short)--exhibits a surprisingly subtle behavior. 
In this letter, we will describe the equilibrium states associated with these 
``point defects'' in the full nonlinear theory. Remarkably, there exists a discrete 
infinity of states (they are ``quantized''); they can therefore be completely 
classified in terms of the surplus angle $\se$ and a quantum number $n$. For each 
$n$ we show that there is a critical value of $\se$, increasing monotonically with 
$n$, beyond which the cone makes contact with itself; if $\se>35.23$ self-contact 
cannot be avoided. The resulting shapes are not unlike the collars--called ruffs--one 
associates with portraits by Rembrandt or Hals. 

To understand the physics of \econes\ we identify the stresses which underpin 
their geometry. We also address the question of stability. What is the ground 
state? Are all equilibria local minima of the bending energy? To answer these 
questions, one must take care not to lose sight of the local constraint of 
isometry. It is reassuring that one can easily built paper models and put the 
results of this analysis to the test.


\begin{figure}
  \includegraphics[width=0.35\textwidth]{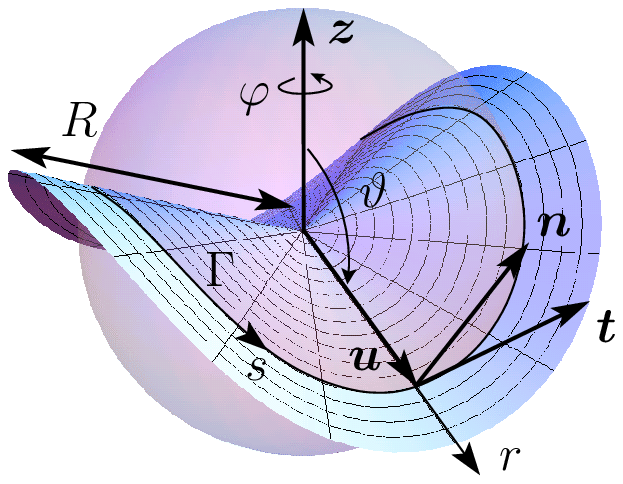}
  \caption[The cone geometry]
  {Geometry of the \econe\ with $\se=\frac{2\pi}{9}$.}
\label{fig:geometry}
\end{figure}

\textit{Geometry and energy.}
The \econe\ can be parametrized in terms of a closed curve $\Gamma:s\to\VECu(s)$
on the unit sphere where the arc-length $s$ runs from 0 to $\st=2\pi+\se$
along this curve (see Fig.~\ref{fig:geometry}). If $r$ denotes the radial 
distance from the origin, the surface is described by the vector function 
$\VECX(r,s)=r\VECu(s)$. Its direction in Euclidean space $\RR^{3}$ will be given 
by the polar and azimuthal angles on the sphere, $\varphi$ and $\vartheta$ 
respectively. The tangent vectors to the \econe\ are $\VECu$ and $\VECt=\VECu'$
where the prime denotes a derivative with respect to $s$. Together with the normal
$\VECn=\VECu\times\VECt$ these vectors form a right-handed surface basis. In the 
direction of $\VECu$ the surface is flat. The curvature along $s$ is given by 
$\kappa=-\VECn\cdot\VECt'$. It is easy to show that $\kappa$ is also the geodesic 
curvature of $\Gamma$ on the unit sphere. It should, however, not be mistaken for 
the Frenet curvature.

Our first task is to identify the configurations that minimize the bending energy 
of an unstretchable cone of radius $R$ with a surplus angle $\se>0$. If we 
introduce the cutoff $r_{0}$ at the apex and integrate over the radial direction,
the bending energy is given by $B = (a/2) \oint_{\Gamma} \romd s \, \kappa^{2}$; 
the dimensional dependence of $B$ is absorbed into the parameter $a=\ln{(R/r_{0})}$.
While it will set the magnitude of the stresses in the cone, it does not play any 
direct role in determining the shape of the \econe. 
The constraint of unstretchability is implemented by adding a term to the energy
functional which fixes the metric via a set of local Lagrange multipliers $T^{ab}$
\cite{paperfolding}. These can be identified with a conserved tangential
stress.


\textit{The shape equation and its solution.}
There is a remarkably simple way to determine the shape of the surface: first 
recall that to every continuous symmetry of a system a conserved Noether current 
exists. As the apex of the \econe\ is fixed, translational invariance is broken. 
The bending energy is, however, rotationally invariant. The corresponding 
conserved vector $\VECJ$, related to the torque about the apex, is given by
\cite{paperfolding}
\begin{equation}
  \VECJ/a = \left( \kappa^{2}/2 - C_{\|} \right) \, \VECn
    + \kappa ' \VECt + \kappa \VECu
  \; .
\end{equation}
We will suppose that $\VECJ$ can be aligned with the $\VECz$ axis, $\VECJ=J\VECz$.
Its square directly yields the first integral of the shape equation of the surface
\begin{equation}
  \tilde{J}^{2} - C_{\|}^{2}
  = \kappa'^{2} + \kappa^{4}/4 + (1 - C_{\|}) \kappa^{2}
  \; ,
  \label{eq:firstintegral}
\end{equation}
where we define $\tilde{J}:=J/a$. The constant $C_{\|}$ is associated with the
fixed arc-length; it will determine the stress established in the surface.
Eqn.~(\ref{eq:firstintegral}) is identical to the equation describing the
behavior of \emph{planar} Euler elastica with (scaled) tension
$\tilde{\sigma}:=C_{\|} - 1$ and $\kappa$ in place of the Frenet curvature.
It is completely integrable in terms of elliptic functions
\cite{LangerSinger,AGM1}.
The \emph{intrinsic} closure condition $\kappa(\st)=\kappa(0)=0$ will provide a
``quantization'' of the solution: the \econe\ has to be periodic in equilibrium 
with a period of $\st/n$, where $n$ is the number of folds. Solving 
Eqn.~(\ref{eq:firstintegral}) for $\kappa$ we obtain
\begin{equation}
  \kappa (s) = 4\sqrt{-k} \;[\ellipticK{k}/S]\;
  \Jacobisnadjustsquareb{2 s \,\ellipticK{k} / S}{k}
  \; ,
  \label{eq:curvature}
\end{equation}
where $S=\st/2n$. The function $\Jacobisnadjustroundb{s}{k}$ is the sine of the Jacobi
amplitude $\Jacobiamadjust{s}{k}$ with parameter $k$. The symbol $\ellipticK{k}$
denotes the complete elliptic integral of the first kind \cite{Abramowitz}.
The parameter $k$ is directly related to the stress $C_{\|}$ and the torque 
$\tilde{J}$. 

To obtain the shape, consider the projections of $\VECJ$ with respect to the
local trihedron. Projecting onto $\VECu$
\begin{equation}
  \tilde{J} (\VECz\cdot\VECu) = \tilde{J} \cos{\vartheta} = \kappa
  \label{eq:proju}
  \;
\end{equation}
yields the polar angle $\vartheta$ as a function of $s$ since $\kappa(s)$ is 
known. Eqn.~(\ref{eq:proju}) places a strong constraint on the equilibrium 
shape. In particular, it implies that the only equilibrium shape consistent 
with a deficit angle is a circular cone. 
Projecting $\VECJ$ onto $\VECt$ reproduces the derivative of 
Eqn.~(\ref{eq:proju}). The remaining projection
\begin{equation}
\tilde{J} (\VECz\cdot\VECn)
    = \tilde{J} \,\varphi' \sin^{2}{\vartheta}
    =  \kappa^{2}/2 - C_{\|} 
  \label{eq:projn}
  \;
\end{equation}
allows one to determine the azimuthal angle $\varphi$ via a simple integration.

The closure of the surface in Euclidean space $\RR^{3}$ provides a second 
(\emph{extrinsic}) closure condition, \ie, $\varphi(\st) = 2 \pi$. This 
equation identifies the constant $k$ implicitly as a function of the surplus 
angle $\se$ and the quantum number $n$. 
Its numerical solution can be approximated remarkably well by a polynomial of
third order in $\se$. For small $\se$, we find $k=a_{1}\se+\CALO(\se^{2})$, 
where $a_{1} = -\frac{1}{2\pi}(1-\frac{1}{n^{2}})$.


\begin{figure}
  \includegraphics[scale=1.0]{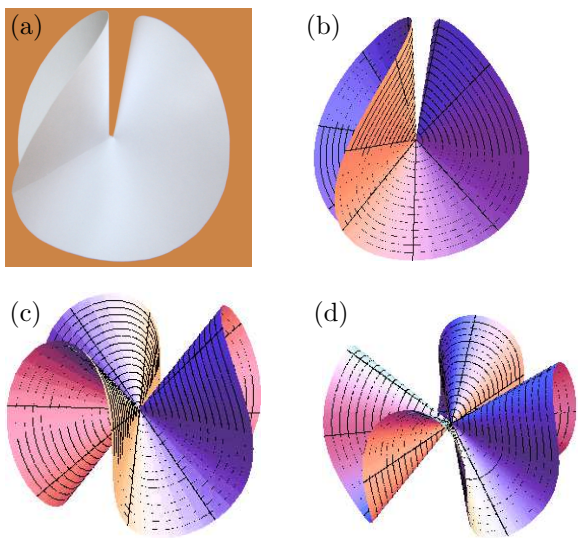}
  \caption[Surface shapes]
  {Paper model (a) and calculated surface shapes for $\se=2\pi$ with $n=2$ (b),
   $n=3$ (c), and $n=4$ (d).}
\label{fig:shapes}
\end{figure}

\textit{Surface shapes.}
The \econe\ has to have two or more folds. This is a consequence of the 
four-vertex theorem \cite{FuchsBook,fourvertexsphere}. If the surplus angle is 
small, one finds a solution with no self-contact for all natural numbers $n\ge 2$. 
As an example, the first three $n$-folds for $\se=2\pi$ are plotted in 
Fig.~\ref{fig:shapes}(b)--(d). It is child's play to construct paper models; 
the $2$-fold illustrated consists of two circular paper discs, each with a 
radial cut, glued together along the opposite sides of the cut. 
The model [see Fig.~\ref{fig:shapes}(a)] closely resembles the calculated shape 
even though the surplus angle in question involves deformations of the flat 
geometry well outside the linear regime.

As $\se$ is increased, the conical geometry becomes more crowded and, at some 
point, the mathematical surface will intersect itself. This happens first with 
the $2$-fold. The physical surface, of course, will not self-intersect. Where 
different regions come into contact, they will experience forces and they will 
deform accordingly \cite{Boueetal}. To determine the critical surplus angle 
$\sek$ above which this happens, consider the opening angle $\alpha=\vartheta(S/2)$ 
at the turning point. It is given by $\alpha = \arccos{(\tilde{J}^{-1} \kappa_{\text{max}})}$ 
with $\kappa_{\text{max}}=4\sqrt{-k}\,\ellipticK{k}/S$. When $\alpha=0$ the 
two-sides of the $2$-fold touch along the $\VECz$ axis and Eqns.~(\ref{eq:proju}) 
and (\ref{eq:projn}) simplify to $\tilde{J}=\kappa_{\text{max}}$ and 
$\tilde{J}^{2}=2C_{\|}$. This implies that $S^{2} = 4 (1-k) \ellipticK{k}^{2}$.
Solving this equation together with the extrinsic closure condition numerically
yields $k\approx-0.28$ and $\sek\approx7.08$.
For \econes\ with more than two folds, the situation is more complicated. 
Adjacent folds touch pairwise at some nonvanishing opening angle. The 
corresponding \emph{kissing conditions} are now $\varphi'(\sk^{j})=0$ and 
$\varphi(\sk^{j})=(2j-1)\pi/2n$ where $j\in\{1,\ldots,2n\}$.
In Tab.~\ref{tab:kissingpoints} the values of $\sek$ of various $n$-folds are
given.
%
\begin{table}
  \begin{tabular}[t]{p{0.75cm}|ccccccc}
     $n$ & 2 & 3 & 4 & 5 & 10 & 50 & $\to\infty$\\ \hline 
     $\sek$ & \ 7.08 \ & \ 13.30 \ & \ 17.78 \ & \ 21.12 \ & \ 29.38 \ & \ 34.92  \ & \ 35.23
  \end{tabular}
  \caption{Kissing points for different $n$-folds.}
  \label{tab:kissingpoints}
\end{table}
%
Interestingly, $\sek$ converges to $\sem\approx 35.23$ from below if $n$ is sent
to infinity. This implies that one cannot find a stable surface with $\se>\sem$ 
which does not make contact with itself. The detailed analysis of these touching 
geometries lies beyond the scope of this letter. The $2$-fold, however, is 
straightforward to study since it will touch itself only at two segments of the 
unit circle of arc-length $\varphi_{t}$ each. Using the kissing conditions with 
$n=2$, $\varphi_{t}$ and $C_{\|}$ can be determined simultaneously for any given 
$\se$.


\begin{figure}
  \includegraphics[width=0.4\textwidth]{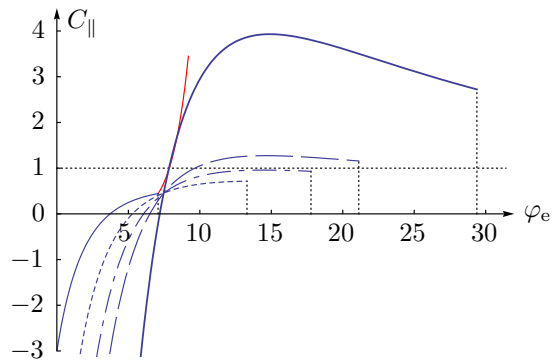}
  \caption[Stresses]
  {Stress $C_{\|}$ as a function of $\se$ for $n=2$ (solid
  line), 3 (short-dashed), 4 (dash-dotted), 5 (long dashed), and
  10 (bold solid). The red curve shows $C_{\|}$ for the touching $2$-fold. 
  Above $C_{\|}=1$ the tension $\tilde{\sigma}$ along the curve $\Gamma$ changes 
  sign.
  }
  \label{fig:stresses}
\end{figure}

\textit{Stresses in the \econe.}
Even though there are no external forces acting on the cone, stresses will be 
set up in the surface due to bending as well as the constraint on the metric. 
The stress tensor $T^{ab}$ which fixes the latter is purely tangential. It is 
diagonal with respect to the basis $(\VECu,\VECt)$ and constant along curves of 
constant radial distance $r$. Its nonvanishing components along $\VECt$ and 
$\VECu$ are given by 
$T_{\|}=-C_{\|}/r^{2}$ and, for a large disc, $T_{\perp}=-T_{\|}$ 
\cite{paperfolding}; it is non-isotropic.

In Fig.~\ref{fig:stresses} $C_{\|}$ is plotted as a function of $\se$ for 
different values of $n$. For small values of $\se$ the expansion 
$k\approx a_{1}\se$ can be used to write 
$C_{\|} = (1-n^{2}) - (4\pi)^{-1} (3 - 7n^{2}) \se + \CALO(\se^{2})$. 
In this regime the stress $C_{\|}$ is negative; this corresponds to a compressive 
stress along the tangential direction; an equal compensating tensile stress will 
act radially. $C_{\|}$ is non-vanishing when $\se =0$, representing the critical 
compression necessary to buckle the planar sheet into the corresponding mode. For 
a fixed surplus angle the absolute value of $C_{\|}$ increases with $n$.
If $\se$ is increased, the curves for different $n$ converge and appear to 
intersect in a single point (the $2$-fold is exceptional making contact with 
itself before this point is reached). Investigated more carefully, however, a set 
of adjacent intersection points is found which converge to $\se\approx 7.47$ for 
$n\to\infty$. Above this region each curve reaches a maximum which diverges 
quadratically with $n$. If $n>5$ and $\se$ sufficiently large, $C_{\|}$ is greater 
than 1. This implies that the tension $\tilde{\sigma}$ in the corresponding planar 
Euler elastica changes sign [\cf text below Eqn.~(\ref{eq:firstintegral})].

However, one must remember that the full stress in the \econe\ includes a 
contribution due to bending. This becomes increasingly important as $\se$ gets 
larger. The tangential projection $\VECf_{\|}$ of the \emph{full} stress tensor 
can be written as $\VECf_{\|}=(\tilde{J}/r^{2}) \sin{\vartheta}\,\VECphi$, where 
$\VECphi$ is the basis vector of $\varphi$. The transmitted force per length 
along $\Gamma$ is a maximum at the equator and a minimum at the turning points. 
It always points in the direction of $\VECphi$; the tangential part of 
$\VECf_{\|}$ becomes tensile where $\varphi'$ is negative. These results can 
easily be verified by cutting the paper model along the flat direction and 
observing how the sheet reacts.


\begin{figure}
  \includegraphics[width=0.4\textwidth]{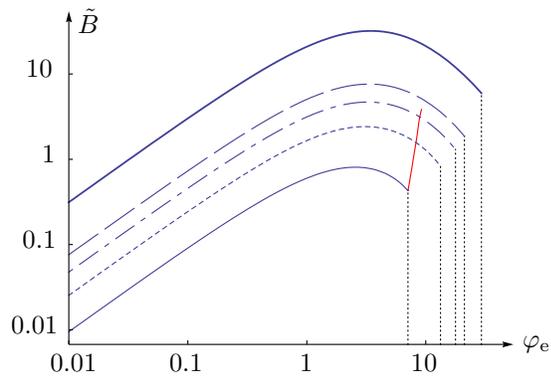}
  \caption[Scaled bending energy]
  {Scaled bending energy $\tilde{B}$ for $n=2$ (solid line), 3 (short-dashed), 
  4 (dash-dotted), 5 (long dashed), and 10 (bold solid). 
  The red curve shows $\tilde{B}$ for the touching $2$-fold. 
  }
\label{fig:bendingenergy}
\end{figure}

\textit{Bending energy.} 
We are also able to obtain an analytical expression for the bending energy 
$\tilde{B}:=B/(a \st /2)$ using the expression for $\kappa$ in 
terms of elliptic functions: 
$\tilde{B} = 64 \ellipticK{k} \, \big[\ellipticcE{k} - \ellipticK{k}\big] 
\, (n  /\st)^{2}$. 
We have normalized the bending energy by dividing by the area of the \econe. The 
function $\ellipticcE{k}$ is the complete elliptic integral of the second kind 
\cite{Abramowitz}.

For fixed $\se$, $\tilde{B}$ scales essentially with $n^{2}$ since $k$ depends 
only weakly on $n$. In Fig.~\ref{fig:bendingenergy} the energy is plotted as a 
function of $\se$ for the lowest $n$-folds. For small $\se$, all curves behave 
as $\tilde{B}\approx\frac{1}{\pi}(n^{2} - 1)\se$. The 2-fold is the ground state 
if $\se$ is below the critical value associated with self-intersection. This 
observation is confirmed nicely by the paper model [see Fig.~\ref{fig:shapes}(a)].
For higher values of $\se$ one needs to examine the folding pattern that is
consistent with self-contact. Initially, by continuity, it will remain the ground 
state. However, with increased crowding one begins to force up the average curvature: 
above $\se=8.27$ the 3-fold possesses lower energy than the touching symmetrical 
2-fold and the \econe\ may flip from $n=2$ to $n=3$. Equivalent behavior is 
expected if $\se$ is increased. 
To analyze the stability of our solutions, it is necessary to examine the second 
variation $\delta^{2}H$ of the total energy functional. This is complicated by 
the fact that the local constraint of isometry has to be imposed on the 
deformations about the conical background. What is remarkable is 
that the calculation is tractable. One can show that $\delta^{2}H$ is of the 
form $\oint \romd s \; \phi \FRAKL \phi$ where $\phi=\VECn\cdot\delta\VECX$ 
denotes the normal deformation of the surface. The relic of 
the isometry constraint is that $\oint \romd s \;\kappa \phi=0$.
The operator $\FRAKL=\partial_{s}^{4}
    + \frac{1}{2} \partial_{s} V_{1} \partial_{s}
    + \frac{1}{2} (V_{2} + \frac{1}{2}V_{1}'')$
is self-adjoint and of fourth order in $\partial_{s}$.
The potentials $V_{1}$ and $V_{2}$ are functions of $\kappa$ and its derivatives.
Using a decomposition of $\phi$ into Fourier modes, one can determine the 
eigenvalues of $\FRAKL$ for arbitrary $\se$ and $n$. They are all positive; 
\econes\ free of self-contacts are stable.


\textit{Conclusions.}
We have described the equilibrium states of a cone exhibiting a surplus angle. 
If we suppose that growth is slow compared to any viscoelastic timescale, 
the surface finds its equilibrium and the approach we have presented describes the 
evolution of the shape of a growing conical tissue. 
If the circumferential arc-length increases linearly with the geodesic radius, 
the surplus angle will remain constant; the cone will scale as it grows. 
Another mode of growth involves an increasing surplus angle. An initially flat
disc will develop into a $2$-fold, although a fluctuation may favor some 
higher energy state which we have seen is stable. However, if at some point, the 
surplus angle reaches $\sek (2)$ the surface will make contact which is costly 
energetically. At some higher value one would expect the surface to flip spontaneously 
into a $3$-fold. This will continue through a $4$-fold and so fifth with 
increasingly higher speed. Above $\sem$, however, self-contact becomes unavoidable. 
Internal local pressure will build up as the spherical volume occupied by the cone 
is packed more and more densely.

What we have learned about the \econe\ lays the foundation for understanding more 
general morphologies. If a disc surrounding the apex is removed, the cone can relax 
into some other flat geometry. Indeed one can easily verify with a paper model that 
the $n=2$ ground state is unstable with respect to such deformations.
These truncated cones can also be glued together to model surfaces which are not 
flat: a surface of constant negative Gaussian curvature can be approximated by a 
telescope formed by such annuli.


\begin{acknowledgments}
Partial support from CONACyT grant 51111 as well as 
DGAPA PAPIIT grant IN119206-3 is acknowledged.
The authors would like to thank A. Boudaoud, L. Bou\'e, and P. V\'azquez 
for helpful discussions.
\end{acknowledgments}


\appendix




\begin{thebibliography}{99}

\bibitem{Jones}
R. A. L. Jones,
\textit{Soft Condensed Matter}
(Oxford University Press, 2002).

\bibitem{RubinsteinColby}
M. Rubinstein and R. H. Colby,
\textit{Polymer Physics}
(Oxford University Press, 2003).

\bibitem{Helfrich}
W. Helfrich,
Z. Naturforsch. \textbf{28c}, 693 (1973).

\bibitem{Seifert97}
U. Seifert,
Adv. Phys. \textbf{46}, 13 (1997).

\bibitem{Wittenreview}
T. A. Witten,
Rev. Mod. Phys. \textbf{79}, 643 (2007).

\bibitem{Sharon}
Y. Klein \etal, 
Science \textbf{315}, 1116 (2007).

\bibitem{JulienMartine}
J. Dervaux and M. Ben Amar, Phys. Rev. Lett. \textbf{101},
068101 (2008).

\bibitem{Witten}
A. Lobhovsky \etal, 
Science \textbf{270}, 1482 (1995).

\bibitem{benpom}
M. Ben Amar and Y. Pomeau, Proc. R. Soc. Lond. A \textbf{453},
729 (1997).

\bibitem{cerdamaha}
E. Cerda and L. Mahadevan, Phys. Rev. Lett. \textbf{80},
2358 (1998).

\bibitem{algae}
K. A. Serikawa and D. F. Mandoli, Planta \textbf{207},
96 (1998).

\bibitem{paperfolding}
J.\,Guven and M.\,M.\,M{\"u}ller, J. Phys. A \textbf{41},
055203 (2008).

\bibitem{LangerSinger} J. Langer and D. A. Singer,
J. Differ. Geom. \textbf{20}, 1 (1984).

\bibitem{AGM1}
J. Arroyo \etal,
J. Phys. A \textbf{39}, 2307 (2006).

\bibitem{Abramowitz}
\textit{Handbook of Mathematical Functions},
9th ed., ed. by M. Abramowitz and I. A. Stegun
(Dover, New York, 1970).

\bibitem{FuchsBook} 
D. Fuchs and S. Tabachnikov,
\textit{Mathematical Omnibus: Thirty Lectures on Classic Mathematics}
(American Mathematical Society, 2007).

\bibitem{fourvertexsphere}
J. J. Nu\~no Ballesteros and M. C. Romero Fuster,
J. Geom. \textbf{46}, 119 (1993).

\bibitem{Boueetal}
L. Bou\'e \etal, Phys. Rev. Lett. \textbf{97}, 166104 (2006).
\end{thebibliography}
\end{document}